\documentclass[useAMS]{mn2e}
\usepackage{times,psfig} 
\begin{document} 
% \hsize=6truein 
 
\title[Are we missing baryons in Galaxy Clusters?]
{Are we missing baryons in Galaxy Clusters?}
 
\author[] 
{\parbox[]{6.in} {S. Ettori \\ 
\footnotesize 
European Southern Observatory, Karl-Schwarzschild-Str. 2, 
D-85748 Garching, Germany \\ }}
\maketitle 
 
\begin{abstract} 
The recent constraints on the cosmological parameters
put from the observations of the WMAP satellite
limit the cosmic baryon fraction in a range that is larger
than, and marginally consistent with,  
what is measured in galaxy clusters. This rises the question
whether or not we are considering all the ingredients of 
cluster baryonic budget. 
Carefully weighing the baryons in X-ray emitting
plasma and stars in cluster galaxies, I conclude that 
the cluster baryonic pie is made by
13 (with a $1 \sigma$ range of 8--19) per cent
of stars, 70 (56--89) per cent of intracluster hot medium
and 17 (0--33) per cent, and a probability of 73 per cent 
of being larger than 0, of ``other'' baryons, presumably 
in the form of warm ($10^5-10^7$ Kelvin) material.

\end{abstract} 
 
 
\section{INTRODUCTION} 
 
The recent analysis of the angular power spectrum 
of the Cosmic Microwave Background (CMB) obtained from WMAP
(Bennett et al. 2003) has provided constraints on the cosmological parameters
(Spergel et al. 2003) that confirms with greater accuracy 
the current energy density of the Universe to be comprised by about 73 per cent
of dark energy and 27 per cent of matter, mostly non-baryonic and dark.
In particular, the quoted constraint on the baryon density,
$\Omega_{\rm b}$, is $0.0224 \pm 0.0009 h_{100}^{-2}$, and on the 
{\it total} matter density, $\Omega_{\rm m}$, is $0.135^{+0.008}_{-0.009} 
h_{100}^{-2}$. Consequently, the cosmic baryon fraction, 
$\Omega_{\rm b} / \Omega_{\rm m}$, is equal to $0.166^{+0.012}_{-0.013}$
and the ratio between baryon and {\it cold} dark matter density,
$\Omega_{\rm c} = \Omega_{\rm m}-\Omega_{\rm b}$, is equal to
$0.199^{+0.017}_{-0.019}$.
These values are expected to be maintained in regions at high 
overdensities that collapse to form galaxy clusters.

The clusters baryon budget is composed mainly from the X-ray 
luminous baryons,  $M_{\rm gas}$, of the intracluster medium (ICM) 
that becomes hotter upon falling
into the cluster dark matter halo by gravitational collapse.
Other contributions come from the baryonic stellar mass 
in galaxies, $M_{\rm gal}$, and from other ``exotic'' sources,
like intergalactic stars and a still poorly defined baryonic dark
matter.
Given the large uncertainties on the relative contribution
from baryons that are not accounted for in either $M_{\rm gas}$
or $M_{\rm gal}$, I qualify these as
``other baryons", $M_{\rm ob}$, as already done in a previous work
(Ettori 2001) in which I discussed the constraints on cluster
baryon budget from BOOMERANG and MAXIMA-I data. The tighter constraints
on the cosmological parameters provided from WMAP allow now more
firm conclusions.

Therefore, one can put the following relation between the relative  
amount of baryons in the Universe and in clusters with total gravitating  
mass, $M_{\rm tot}$:
\begin{equation} 
\frac{\Omega_{\rm b} }{\Omega_{\rm m}} = \frac{M_{\rm b}}{M_{\rm tot}} = 
\frac{f_{\rm gas}}{Y\, B\, C} + \frac{f_{\rm gal}}{B} + \frac{f_{\rm ob}}{B},
\label{eq:fbar} 
\end{equation} 
where $f_{\rm gas} = M_{\rm gas}/M_{\rm tot}$,  
$f_{\rm gal} = M_{\rm gal}/M_{\rm tot}$ ($\approx 
0.010^{+0.005}_{-0.004} h_{100}^{-1}$ in Fukugita et al. 1998),  
$f_{\rm ob} = M_{\rm ob}/M_{\rm tot}$, and
$Y$ is the parameter representing the cosmic depletion of baryons
at the virial radius with respect to the global value 
($\approx 0.92 \pm 0.06$ from the hydrodynamical simulations of 
the Santa Barbara Project; Frenk et al. 1999).
I parametrize the uncertainties on the measurements of the 
total gravitating mass and gas mass through the factors 
$B$ and $C$, respectively.
These factors act to increase the total mass
estimates (i.e. $B>1$) if corrections to the hydrostatic
equilibrium equation are required for bulk motions of the ICM 
or non-thermal pressure support, and to lower the true gas mass (i.e. $C>1$)
if clumpiness is present in the ICM that is assumed to be smoothly distributed 
(e.g. Mathiesen et al. 1999). 

% It is worth noticing that $M_{\rm tot}$ as inferred from X-ray analysis
% through the equation of the hydrostatic equilibrium between the gas and 
% the gravitational potential is, strictly speaking, just {\it all the 
% gravitating mass that is not in the X-ray gas}. 
% In fact, hydrostatic equilibrium
% holds between the X-ray emitting gas and the {\it external} gravitational
% force that supports it, and the equation generally adopted is a 
% first-order differential equation that does not include the self-gravity 
% of the gas.
% Therefore I define hereafter $M_{\rm tot}$ equal to the sum 
% of the gravitating mass measured in X-ray analysis and $M_{\rm gas}$.
% Another way to proceed is to calculate $\Omega_{\rm b}/\Omega_{\rm c}$ 
% and compare it to the baryon fraction estimated by the gravitating mass 
% measured in X-ray analysis, subtracted of $M_{\rm gal} + M_{\rm ob}$.

In this Letter, I will analyze the equation~\ref{eq:fbar} (i) to assess the
consistency between the cosmic and the cluster baryon budget
and (ii) to put significant constraints on $f_{\rm ob}$.
In a consistent way, I adopt the WMAP results on the Hubble constant, $H_0$,
of $71^{+0.04}_{-0.03}$ km s$^{-1}$ Mpc$^{-1}$ to rescale 
all the measured quantities.

\begin{figure*} 
\hspace*{.6cm} \hbox{
  \psfig{figure=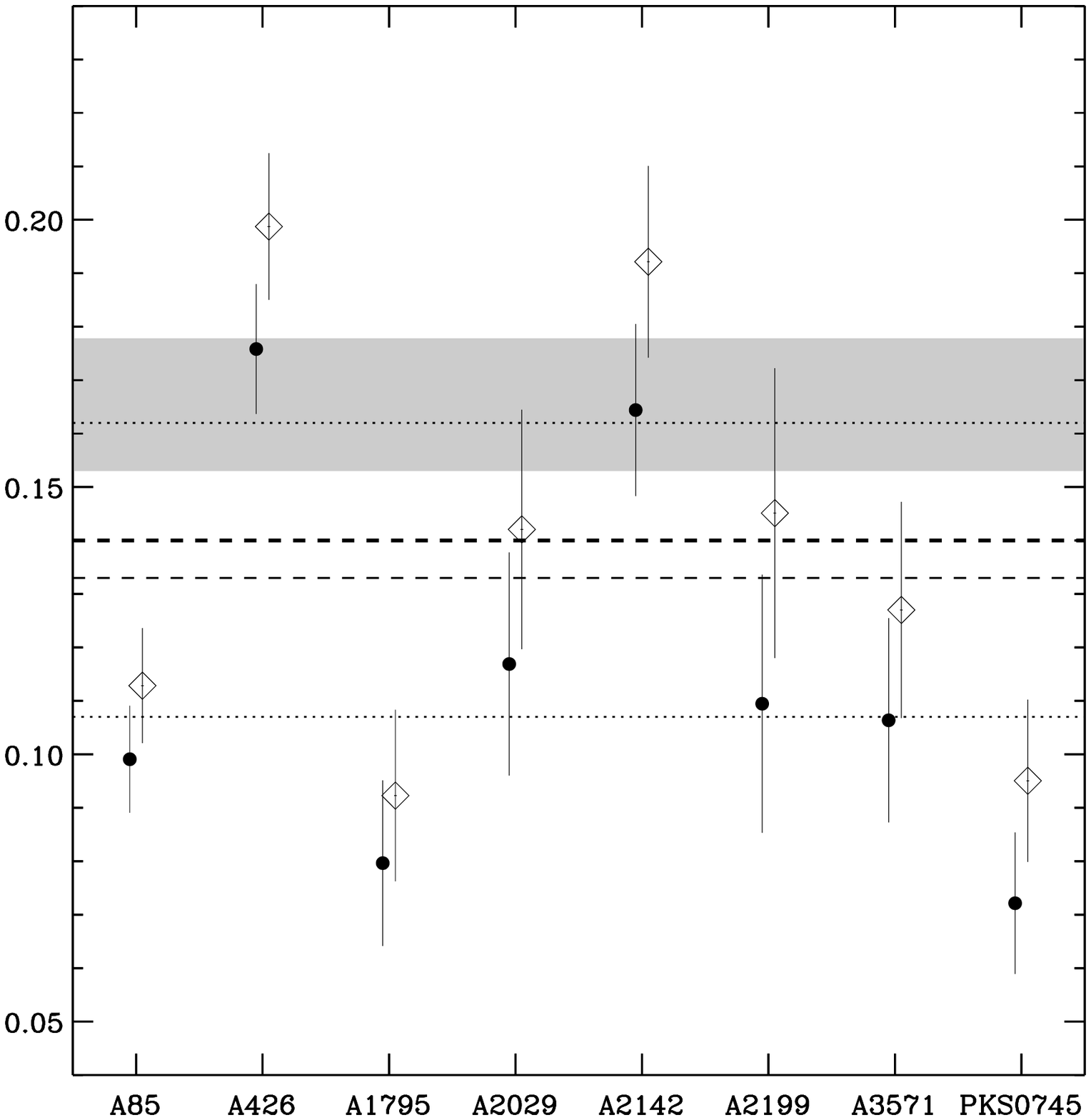,width=0.45\textwidth,angle=0} 
  \psfig{figure=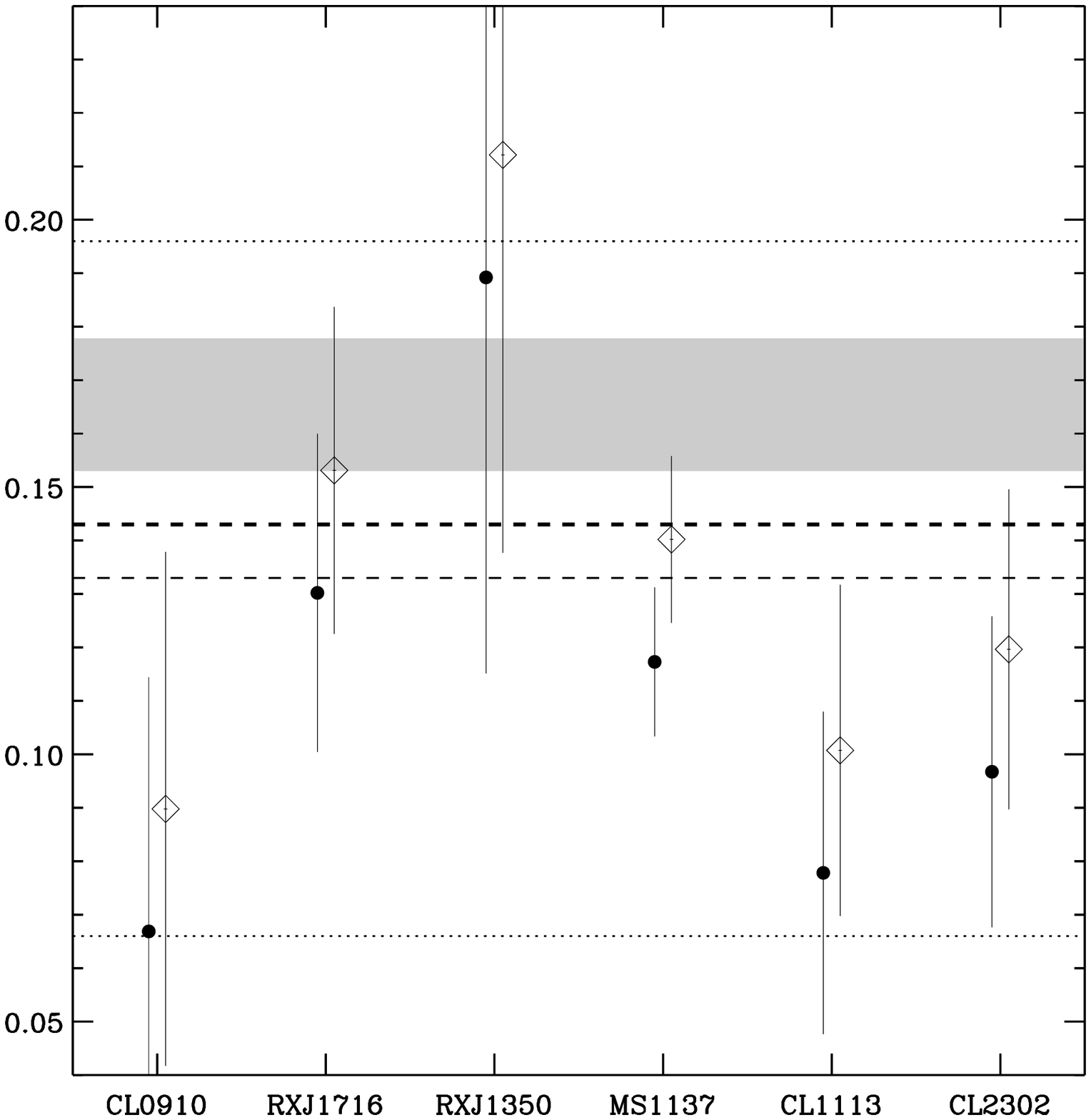,width=0.45\textwidth,angle=0} 
}
\caption[]{Gas ({\it dots}) and baryon ({\it diamonds}) fractions
estimated at an overdensity of 200
for a cosmology fixed to the best-fit WMAP results (Spergel et al. 2003).
These results constrain the cosmic baryon budget to the shaded region. 
The central ($\pm 1 \sigma$) value of the observed distribution
is obtained with a Bayesian method (see Press 1996) and is indicated
with a {\it dashed} ({\it dotted}) line. The thickest {\it dashed} line
indicates the estimate of the baryon fraction corrected by the depeletion
parameter $Y = 0.92$.
{\it (Left)} Low ($z<0.1$) redshift sample of galaxy clusters
observed with {\it BeppoSAX}. The best-fit spectral results
were deprojected to recover the gas density and temperature 
profile and to infer the gas and gravitating mass (see Ettori,
De Grandi, Molendi 2002). 
The $f_{\rm gal}$ is estimated from B-band luminosities 
in Girardi et al. (2002), assuming $M_{\rm gal}/L_B = 4.5 \pm 1.0
h_{100} M_{\odot}/L_{\odot}$ (Fukugita et al. 1998). 
Where $L_B$ is not available, 
the median value of measured $f_{\rm gal} = 0.022$ obtained 
from six nearby systems is used [Lin et al. (2003, Fig.~7)  
measure $f_{\rm gal}$ between 0.01 and 0.02 from K-band luminosity].
{\it (Right)} High ($z>0.7$) redshift sample of objects
observed with {\it Chandra} (see Ettori, Tozzi, Rosati 2003). 
} \label{fig:dat} \end{figure*} 

\begin{figure*}
\hspace*{-.3cm} \hbox{
  \psfig{figure=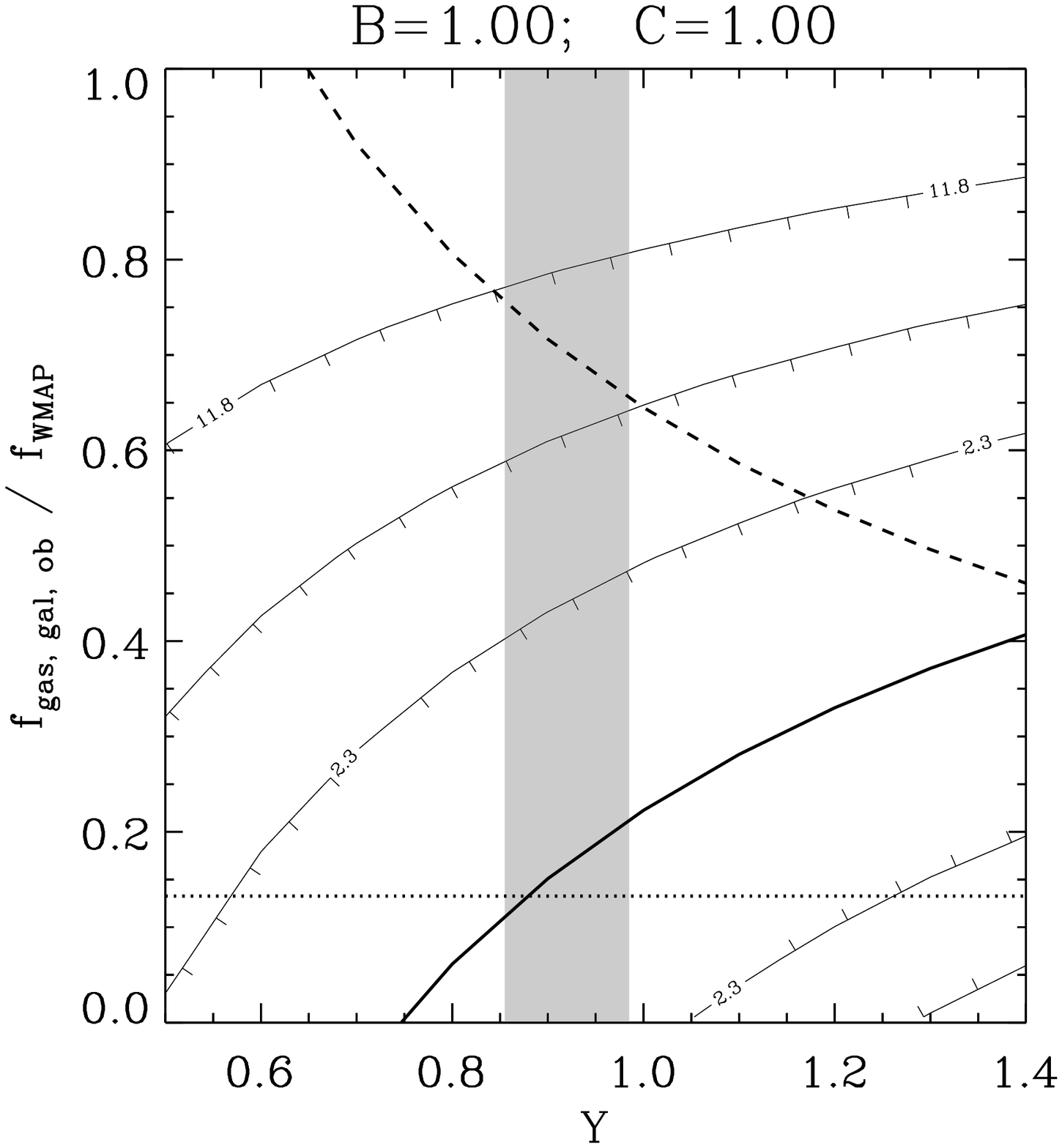,width=0.35\textwidth}
  \psfig{figure=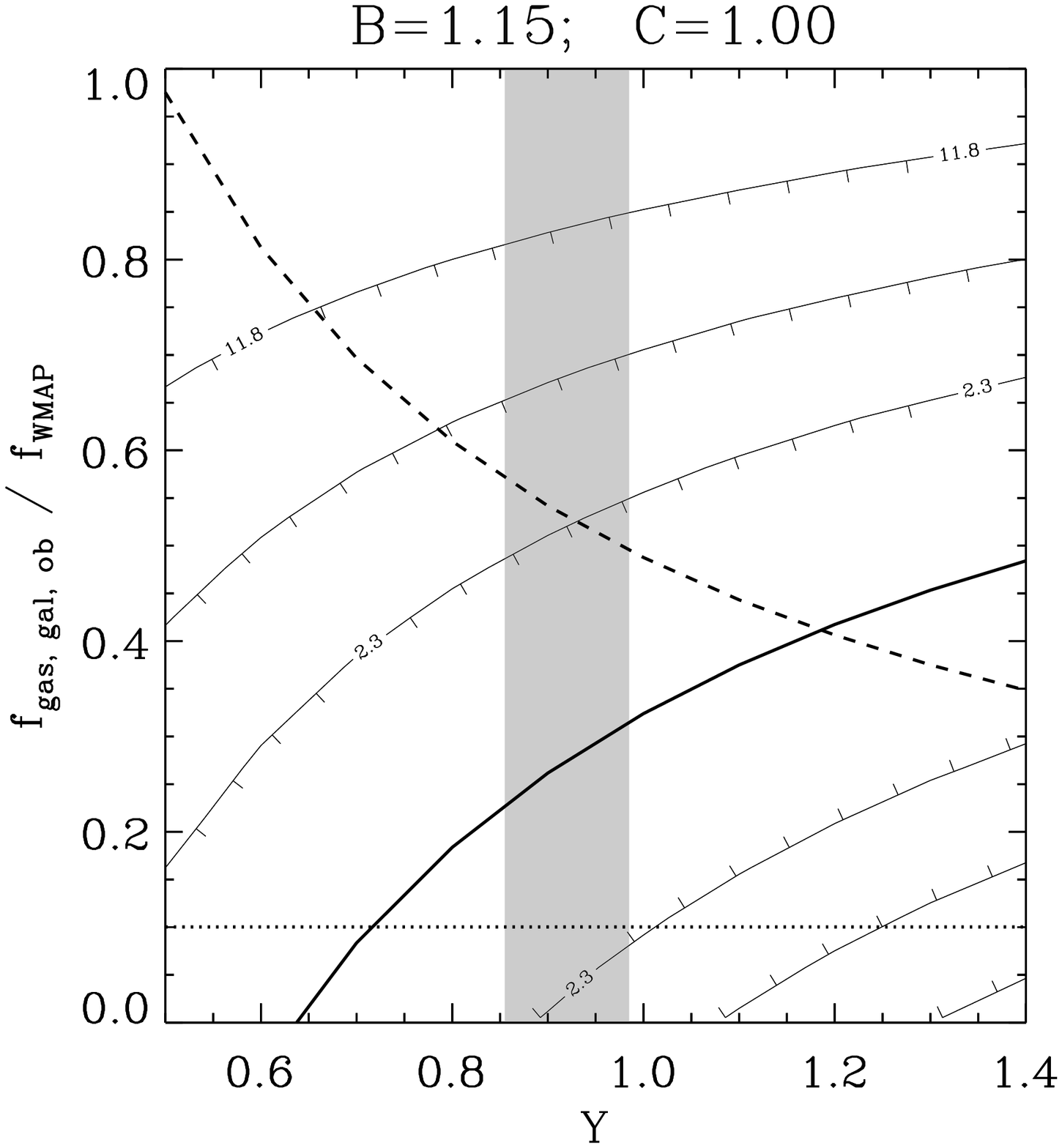,width=0.35\textwidth}
  \psfig{figure=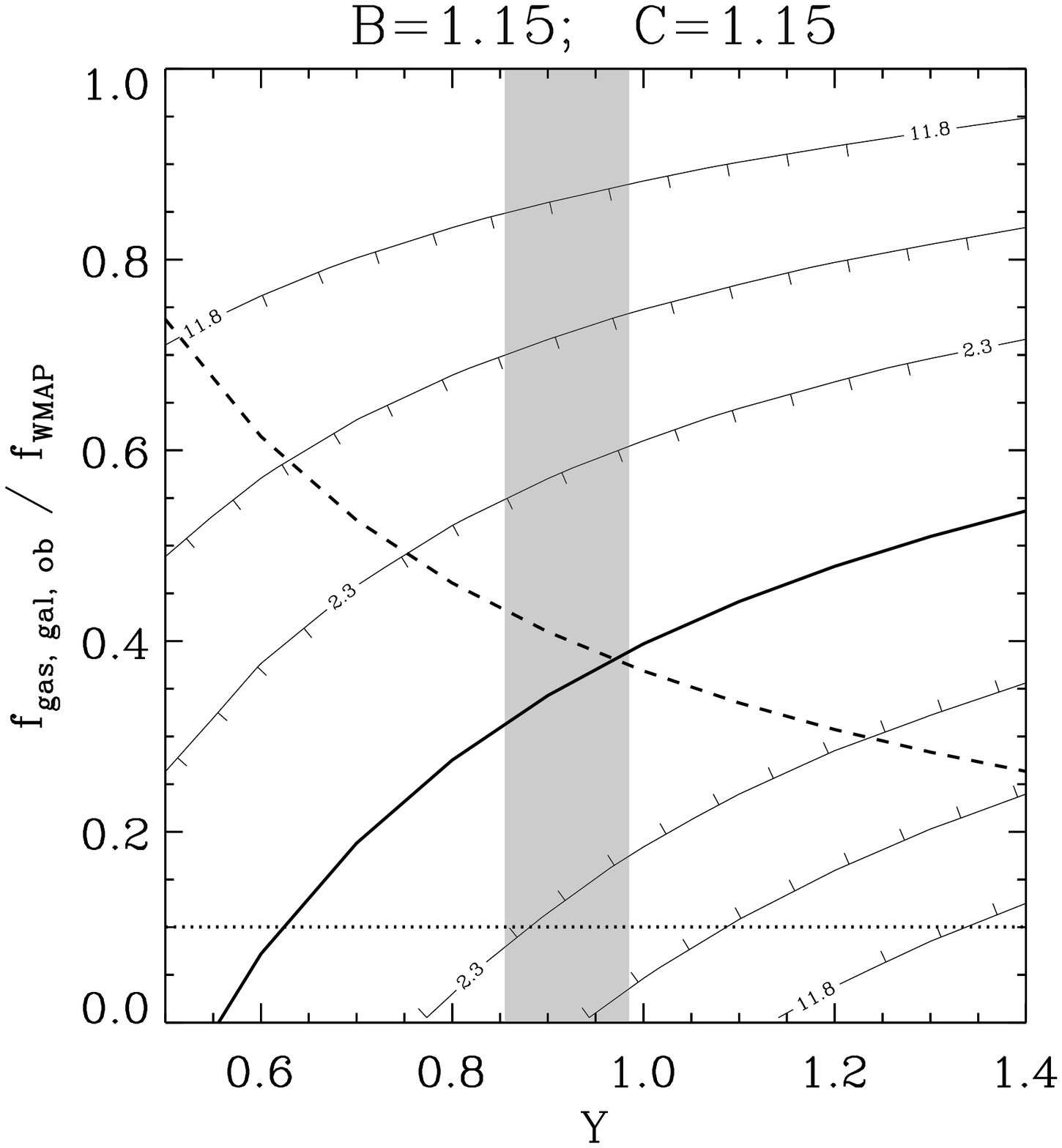,width=0.35\textwidth}
}
\caption[]{Constraints as function of the depletion parameter $Y$
on the gas ({\it dashed line}), stellar ({\it dotted
line}) and ``other baryons" ({\it solid line} and probability contours
at $1, 2, 3$ sigma confidence level) mass fraction normalized to the 
cosmic value. The shaded region indicates the range of $Y$ permitted  
from hydrodynamical simulations (Frenk et al. 1999).
} \label{fig:fob} \end{figure*}

\section{The cosmic baryon budget from WMAP}

Figure~\ref{fig:dat} shows the allowed $1 \sigma$ region
from the WMAP results on the cosmic baryon budget with respect
to the observed gas and baryon fraction for a sample
of relaxed galaxy clusters both at low and high redshift.
All these values are estimated at the overdensity of 200 
with respect to the critical density (i.e. within the cluster
region that numerical simulations show to be virialized in a 
$\Omega_{\rm m}$ independent way; see, e.g., Evrard et al. 2002). 
Using a Bayesian approach (Press 1996), I measure 
the average gas (baryon) fraction and confirm that is consistent between
the two samples: $0.107^{+0.028}_{-0.019} (0.133^{+0.029}_{-0.026})$
and $0.111^{+0.069}_{-0.063} (0.133^{+0.063}_{-0.067})$
(the error-weighted means of the gas fraction distribution are 
$0.116 \pm 0.005$ and $0.111 \pm 0.010$ in the low$-z$ and high$-z$
sample, respectively. Note that these values do not include 
any correction by the baryonic depletion that is expected to raise
the gas/baryon fraction by about 8/6 per cent at this overdensity.
See also Fig.~\ref{fig:dat} and text that follows).
These estimates are consistent with other, independent, recent
determinations (e.g. Allen et al. 2001, Pratt \& Arnaud 2002) and
consistently lower than the baryon budget required from WMAP results.
It is worth noticing that only the highest estimates of the
gas (baryon) mass fraction (e.g. A426, A2142, RXJ1350)
are perfectly consistent with WMAP results, whereas the other clusters
are systematically below them.

To investigate the systematics that could affect this estimate,
I change the values of the factors $B$ and $C$ and 
study their influence on the baryon fraction.
The factor $B$, which
parametrizes the uncertainties on $M_{\rm tot}$ is expected
to be between 1 and 1.15 from the cluster mass profiles recovered
from both X-ray and lensing data (e.g. Allen et al. 2001).
The factor $C$ represents the level of clumpiness that 
affects the estimate of $M_{\rm gas}$ in X-ray analysis
and that simulations show to be lower than 1.2 (Mathiesen et al. 1999).

Figure~\ref{fig:fob} shows that higher values of $B$ and $C$
require a more relevant role to be played by $f_{\rm ob}$, 
giving a $2 \sigma$ positive detection for typical value of $Y$
when $B$ and $C$ are 15 per cent larger than the null hypothesis
of reliable estimates of both $M_{\rm gas}$ and $M_{\rm tot}$
from X-ray analysis.   

\begin{figure*}
% \hbox{
%   \psfig{figure=rgal_100_100.ps,width=0.5\textwidth}
%   \psfig{figure=rgal_115_115.ps,width=0.5\textwidth}
% } 
\vspace*{-0.4cm}
\hspace*{0.6cm} \hbox{
  \psfig{figure=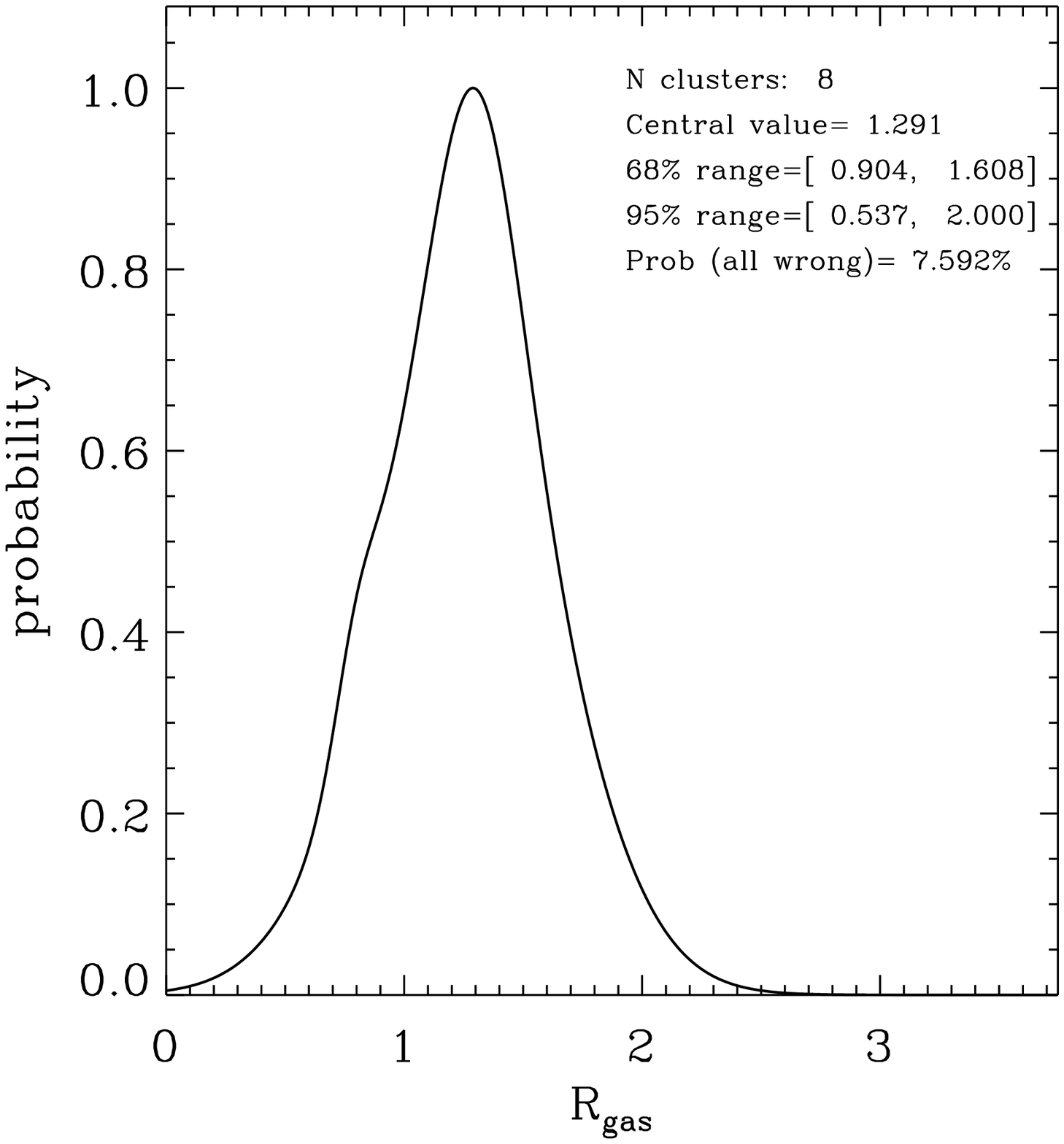,width=0.45\textwidth}
  \psfig{figure=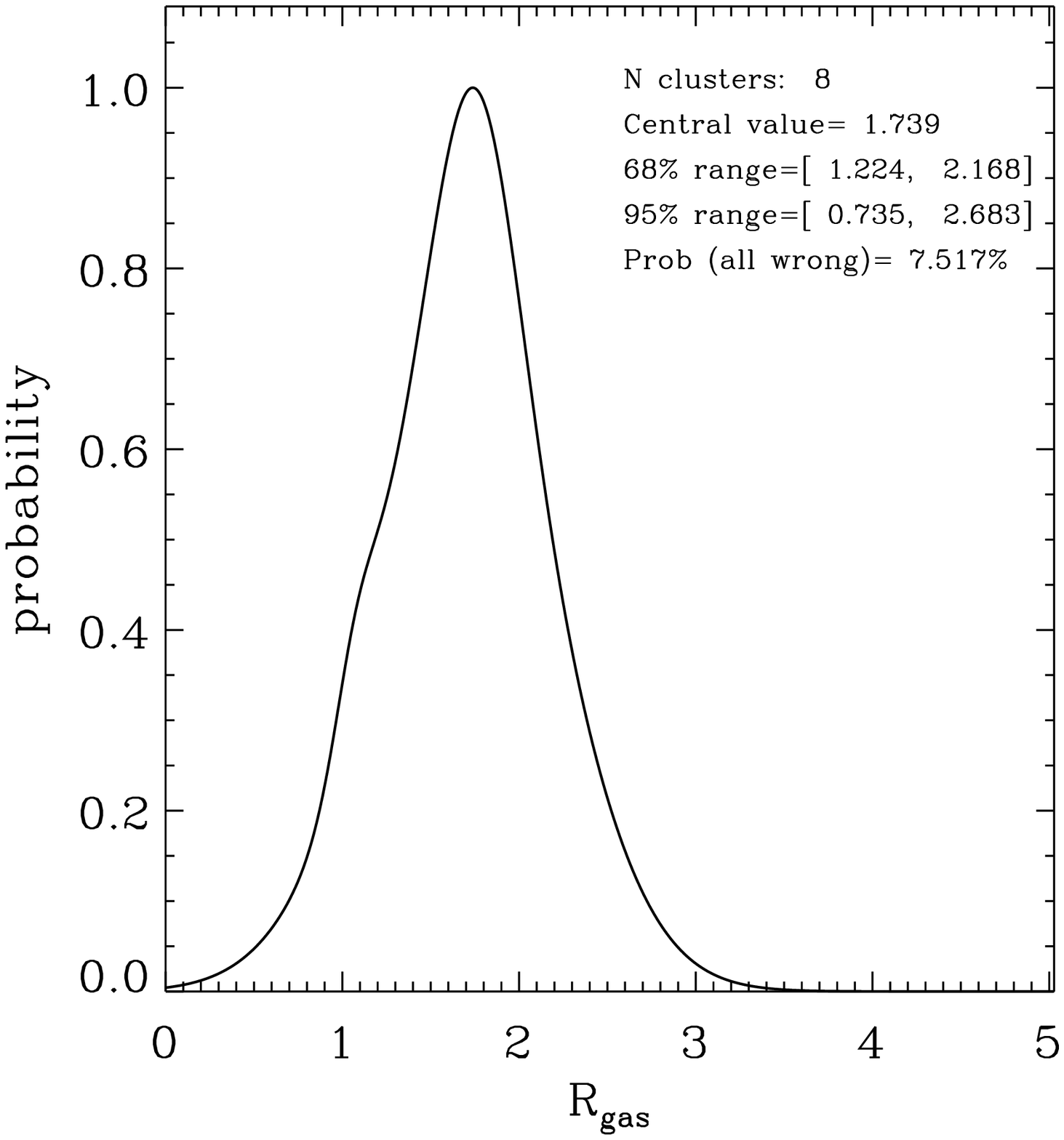,width=0.45\textwidth}
} \caption[]{Bayesian distribution (Press 1996) of the ratio $R_{\rm gas} = 
(f_{\rm WMAP}-f_{\rm gal})/f_{\rm gas}$ for $B=C=1$ (left) and
$B=C=1.15$ (right). 
} \label{fig:macho} \end{figure*}

Moreover, 
I can estimate from the observables the ratio 
$R_{\rm gas} = (\Omega_{\rm b}/\Omega_{\rm m} -f_{\rm gal}/B)
/ f_{\rm gas} \times (Y B C) \approx (f_{\rm ob}/f_{\rm gas} +1)$ 
and to evaluate the probability that 
$R_{\rm gas} > 1$ and a no-zero value of $f_{\rm ob}$  
is required from the data (see Fig.~\ref{fig:macho}).
I obtain values of $R_{\rm gas}$ between 1.3 ($B=1, C=1$)
and 1.7 ($B=1.15, C=1.15$),
with an interval accepted at the 95 per cent confidence level
of 0.5--2.7.
More significantly, this ratio has to be larger than 1 
at 76.6 ($B=1, C=1$) and 92.5 ($B=1.15, C=1.15$) per cent confidence level. 
This result gives a high confidence to the conclusion that a 
significant amount of baryons has to be present apart from what 
is observed both at X-ray and optical wavelength. 
 
\begin{figure}
\psfig{figure=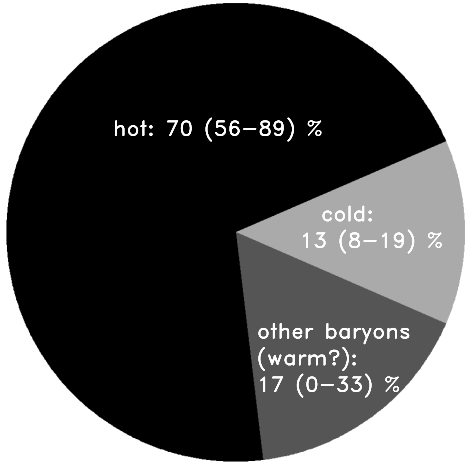,width=0.4\textwidth}
\caption[]{Cluster baryonic pie. The $1 \sigma$ range from the 
Bayesian calculations applied to the eight nearby clusters
is shown in parenthesis. A depletion factor $Y = 0.92 \pm 0.06$ 
is assumed.
The factors $B$ and $C$ are fixed to 1. We expect larger values of the 
{\it warm} ingredient if either $B$ or $C$ are $>1$.
It is worth noticing that the {\it hot} component, $f_{\rm gas}$, 
scales as $h^{-1.5}$ and the {\it cold} component, $f_{\rm gal}$, is
here independent from $h$.
} \label{fig:pie} \end{figure}

\section{Conclusions} 
 
By comparing the recent cosmological constraints from measurements 
of the angular power spectrum of the temperature anisotropy 
in the CMB done with WMAP with the observed distribution of the gas mass 
fraction in clusters of galaxies, I conclude that 
 
\begin{enumerate} 
\item galaxy clusters with the highest observed gas/baryon fraction are
well in agreement with the WMAP estimate of the cosmic 
baryon budget.
On average, however, a disagreement in the order of 15--20 per cent
is present, the {\it observed} cluster baryon fraction being a lower estimate   
of the cosmic one. This implies that estimates of the cold
dark matter density, $\Omega_{\rm c}$, done by applying 
mean results from large sample of objects tend to overestimate it.
In this perspective, a more ``realistic" result is provided from
the highest estimate in the distribution of the cluster gas (baryon) mass
fraction; 

\item this dark baryonic component appears to be 
between a fraction, and up to 2 times, the measured gas fraction.
Values of $R_{\rm gas} = (f_{\rm WMAP} - f_{\rm gal})/f_{\rm gas}$
larger than 1 are required from the data with a level of confidence
of about 80 per cent and more, if we are underestimating
(overestimating) the total (gas) mass.
The most probable values are $f_{\rm ob}/f_{\rm gas} =
R_{\rm gas} -1 = $ 0.3 ($B=C=1$) and 0.7 ($B= C=1.15$),
and lower than 1.7 at 95 per cent confidence level.
\end{enumerate}  

It is very  unlikely that galactic objects, such as
``colored" (red, brown, white, beige) dwarfs, stellar remnants and 
other species of MACHOs (see review in Gilmore 1999
and Evans 2003), or intergalactic ones formed from tidal 
disruption of cluster dwarfs, like planetary nebulae 
(e.g. Ciardullo et al. 2002), red-giant-branch 
stars (e.g. Ferguson et al. 1998) and supernovae 
(Gal-Yam et al. 2003), can be responsible for such 
amount of baryons. It is reasonable 
to believe that they can contribute by 
about 0.2 times $f_{\rm gal}$, or $0.04 f_{\rm gas}$.

The most plausible suspect to give so large contribution is then
a X-ray warm ($10^5$ K $< T <$ $10^7$ K) intracluster medium (W-ICM).
Large scale cosmological and hydrodynamical simulations by
Cen \& Ostriker (1999) and Dav\'e et al. (2001) 
show that the mass fraction at redshift 0
is largely dominated from a warm medium, with a relative contribution
in mass that is about 2 times larger than the amount of hot
($T > 10^7$ K) baryons.
However, less than 30 per cent of it 
falls in overdensities $\ge 60$ that are typical for bound
structures in a $\Lambda$CDM universe.
Furthermore, Bonamente et al. (2002)
present evidence of excess in the soft X-ray emission 
between 0.2--0.4 keV in 50 per cent of the 38 clusters
in their sample of {\it Rosat} PSPC observations.
They list several suggestions on how to explain this excess, 
originally observed in extreme ultraviolet (Lieu et al. 1996),
both as thermal and non-thermal component.
If we assign this emission to the baryons that are lacking in our budget, 
we interpret it as thermal emission due either to the diffuse/halo
component of unresolved X-ray faint cluster galaxies or W-ICM. 
In the first case, we are forced to consider an inexplicable large amount 
of X-ray emitting member galaxies. More plausible is then
the hypothesis of W-ICM, even though its cooling time 
tends to be very short with the bulk of the radiation in emission line
if this gas is not primordial.  Fabian (1997) suggested that it can
be located in turbulent mixing layers lying between embedded 
cold clouds and the ICM. However, the traditional
picture on the efficiency of cooling processes in the cluster cores is 
not supported anymore after that XMM and Chandra observations did not
report evidence of gas cooler than 1--2 keV (e.g. Peterson et al. 2003)
and showed a strong interplay between ICM, the central active galaxy 
(e.g. Fabian 2002) and merging 
cool clumps (e.g. Markevitch et al. 2000, Mazzotta et al. 2003).
On the other side, the production of thermal energy per particle
due to supernovae related to the star formation activity is in the 
order of $0.4 (\eta/0.1) (N_{\rm SNII}/10^9)
(10^{13} M_{\odot} / M_{\rm gas})$ keV for a given
efficiency $\eta$ in converting the kinetic energy
of the explosion into thermal energy through galactic winds, 
and adopting typical values of the cluster gas mass and number of 
type II supernovae as required from the observed ICM metallicity.
While this energy per gas particle 
is not enough to stop cooling the hot ICM, it can easily 
accommodate for the survival of the warm component.
  
Intriguingly, the stronger soft excess detected in {\it Rosat}
data from Bonamente et al. (2002) is measured in objects like
A85 and A1795, that are the ones lacking most of the
baryons with respect to the cosmic budget as plotted 
in Fig.~\ref{fig:dat}. Significant detection is also present in
A2029, A2199 and A3571, whereas a marginal detection is associated
to A2142.
Nevalainen et al. (2003; see also Kaastra et al. 2003) 
confirm with XMM the excess in 
the soft X-ray emission in A1795 and that this excess is best fitted by 
a thermal component with a characteristic temperature of 
0.8 keV, which is about an order of magnitude higher than
what required from {\it Rosat} data but still consistent 
with our energetic arguments. Using their estimation
of the atom density of the W-ICM in the core of A1795, and assuming
that it is broadly distributed like the ICM, 
one can infer a $f_{\rm W-ICM} / f_{\rm gas} =
R_{\rm gas} -1 \approx 0.43$. In general, values of 
$f_{\rm W-ICM} / f_{\rm gas}$ between 0.1 and 0.5 are expected.

To summarize, clusters seem to have similar behavior in accumulating
the same relative amount of baryons. It is then their peculiar
thermal history due to the interplay of merging actions and/or 
activity of the central active galaxy that provides the baryonic
ingredients and cook the baryonic pie that we taste and show in
Figure~\ref{fig:pie}.
To prepare it, I have considered only the eight nearby clusters which
provide a more reliable estimate of $f_{\rm gal}$ and are less affected
from systematics in the determination of $f_{\rm gas}$ (see discussion in
Ettori et al. 2003). I have also corrected the gas fraction by the depletion
factor $Y \approx 0.92$.
The baryonic pie is then made of 70 per cent of {\it hot} ICM, 
with $1 \sigma$ range between 56 and 89 per cent and a distribution 
of the calculated $f_{\rm gas}/f_{\rm WMAP}$ that spans between 28 and 143 per 
cent at $2 \sigma$ confidence level (higher upper limit observed in A426 that 
has a most probable $f_{\rm gas}/f_{\rm WMAP}$  value of 115 per cent).
The {\it cold}, stellar component is responsible for 13 ($1 \sigma$: 
8--19) per cent, with an observed distribution in the sample 
between 2 and 37 per cent ($2 \sigma$ lower and upper limit, with the latter
reached in A2199, which has  a central value of 21 per cent).
Finally, a third ingredient, probably a {\it warm} ICM, contributes by about
17 (0--33) per cent (and a probability to be larger than 0 of 73 per cent)
with a distribution that goes from --29$\pm$15 per cent in A426 to
40$\pm$12 per cent in A1795, one of the objects with the largest detected
soft excess (Bonamente et al. 2002, Kaastra et al. 2003). 

\section*{ACKNOWLEDGEMENTS}  
I am grateful to Hans B\"ohringer and Andy Fabian that 
pointed out a relevant weakness in the assumptions done in the 
original manuscript.

\end{document}